\documentclass[aps,twocolumn,showpacs,showkeys]{revtex4}
\usepackage{bm}
\usepackage{graphicx}
\usepackage{epsfig}

\begin{document}

\title{Spin orientation and spin currents induced by linearly polarized
light}
\author{Sergey~A.~Tarasenko, Eugeniyus~L.~Ivchenko}
\affiliation{A.F.~Ioffe Physico-Technical Institute, Russian
Academy of Sciences, 194021 St.~Petersburg, Russia}
\begin{abstract}
To date, optical orientation of free-carrier spins and spin
currents have been achieved by circularly polarized light, while
the linearly polarized light has been used for optical alignment
of electron momenta. Here we show that, in low-dimensional
structures, absorption of the linearly polarized light also leads
to the spin polarization and spin photocurrent, and, thus, the
electron and hole spins can be manipulated by light of zero
helicity. The microscopic description of the both effects is
developed for interband optical transitions in undoped quantum
wells (QWs) as well as for direct intersubband and indirect
intrasubband (Drude-like) transitions in $n$-doped QW structures.
\end{abstract}

%%% PACS numbers


\maketitle

\section{Pure spin photocurrents}

Pure spin current represents a non-equilibrium distribution where
free carriers, electrons or holes, with the spin ``up'' propagate
mainly in one direction and equal number of spin-down carriers
propagates in the opposite direction. This state is characterized
by zero charge current because electric currents contributed by
spin-up and spin-down quasiparticles cancel each other, but leads
to accumulation of the opposite spins at the opposite edges of the
sample. Spin currents in semiconductors can be driven by an
electric field acting on unpolarized free carriers (the so-called
spin Hall effect). They can be induced as well by optical means
under interband or intraband optical transitions in
non-centrosymmetrical bulk and low-dimensional
semiconductors~\cite{Bhat,TI,Zhao,Ganichev}.

The appearance of a pure spin current in semiconductor quantum
wells (QWs) under interband optical pumping with linearly
polarized light is linked with the spin splitting of the energy
spectrum, which is linear in the wave vector ${\bf k}$, and the
spin-sensitive selection rules for the optical transitions. The
effect is most easily conceivable for direct transitions between
the heavy-hole valence subband $hh1$ and conduction subband $e1$
in QWs of the C$_s$ point symmetry, e.g., in (110)-grown QWs. In
such structures the spin component along the QW normal $z'\|[110]$
is coupled with the in-plane electron wave vector. This leads to
$\mathbf{k}$-linear spin-orbit splitting of the energy spectrum as
sketched in Fig.~1, where the heavy hole subband $hh1$ is split
into two spin branches $\pm 3/2$. In the reduced-symmetry
structures, the spin splitting of the conduction subband is
usually smaller than that of the valence band and not shown for
simplicity. Due to the selection rules the allowed direct optical
transitions from the valence subband $hh1$ to the conduction
subband $e1$ are $|+3/2 \rangle \rightarrow |+1/2 \rangle$ and
$|-3/2 \rangle \rightarrow |-1/2 \rangle$, as illustrated in
Fig.~1 by vertical lines. In the presence of the spin splitting,
electrons with the spins $\pm 1/2$ are photoexcited in the
opposite points of the ${\bf k}$ space which results in a flow of
electrons within each spin branch. The corresponding fluxes ${\bf
j}_{+ 1/2}$ and ${\bf j}_{- 1/2}$ are of equal strengths but of
opposite directions. Thus, this non-equilibrium electron
distribution is characterized by the nonzero spin current ${\bf
j}_{\rm spin}$ = $(1/2) ({\bf j}_{+1/2} - {\bf j}_{- 1/2})$ but a
vanishing charge current, $e ({\bf j}_{+1/2} + {\bf j}_{-
1/2})=0$.

\begin{figure}
  \includegraphics[height=.23\textheight]{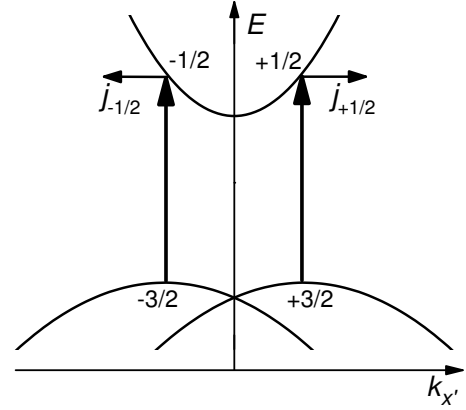}
  \caption{Microscopic origin of pure spin photocurrent caused by spin splitting
  of the band structure.}
\end{figure}

In general, the flux of electron spins can be characterized by a
pseudo-tensor ${\bf J}$ with the components $J_{\beta}^{\alpha}$
describing the flow in the $\beta$ direction of spins oriented
along $\alpha$, with $\alpha$ and $\beta$ being the Cartesian
coordinates. The non-zero components of the photo-induced spin
current are determined by the light polarization and the explicit
form of spin-orbit interaction. The latter is governed by the QW
symmetry and can be varied. In (110)-grown QWs the absorption of
unpolarized light leads to a flow along $x'\parallel[1\bar{1}0]$
of spins oriented along $z'$. This component can be estimated as
\begin{equation}
J_{x'}^{z'} = \gamma_{z'x'}^{(hh1)} \frac{\tau_e}{2 \hbar}
\frac{m_h}{m_e+m_h} \frac{\eta_{cv}}{\hbar\omega} I \:,
\end{equation}
where $\gamma_{z'x'}^{(hh1)}$ is a constant describing the ${\bf
k}$-linear spin-orbit splitting of the $hh1$ subband, $\tau_e$ is
the relaxation time of the spin current, $m_e$ and $m_h$ are the
electron and hole effective masses in the QW plane, respectively,
$\eta_{cv}$ is the light absorbance, and $I$ is the light
intensity.

Another contribution to spin photocurrents may come from ${\bf
k}$-linear terms in the matrix elements of the interband optical
transitions. Taking into account ${\bf k}\cdot{\bf p}$ admixture
of the remote condution band $\Gamma_{15}$ to the valence-band and
conduction-band states $X_{{\bf k}},Y_{{\bf k}},Z_{{\bf k}}$ and
$iS_{{\bf k}}$, one derives the interband matrix elements of the
velocity operator for bulk zinc-blende-lattice
semiconductors~\cite{Khurgin}
\begin{eqnarray}
\langle iS_{{\bf k}} | {\bf e}\cdot{\bf v}| X_{{\bf k}} \rangle =
(P/\hbar) [e_x + i\beta(e_y k_z + e_z k_y)] \:,\\
\langle iS_{{\bf k}} | {\bf e}\cdot{\bf v}| Y_{{\bf k}} \rangle =
(P/\hbar) [e_y + i\beta(e_x k_z + e_z k_x)] \:,\nonumber \\
\langle iS_{{\bf k}} | {\bf e}\cdot{\bf v}| Z_{{\bf k}} \rangle =
(P/\hbar) [e_z + i\beta(e_x k_y + e_y k_x)] \nonumber \:,
\end{eqnarray}
where $\beta=Q P'(2E'_g+E_g)/[PE'_g(E'_g+E_g)]$ is a material
parameter, $P$, $P'$ and $Q$ are the interband matrix elements of
the momentum operator at the $\Gamma$ point multiplied by
$\hbar/m_0$ ($m_0$ is the free electron mass), $E_g$ and $E'_g$
are the energy band gaps, and $x\|[100]$, $y\|[010]$, $z\|[001]$.
For GaAs band parameters~\cite{Jancu} the coefficient $\beta$ can
be estimated as $0.2$~\AA. Calculation shows that, in (110)-grown
QWs, the spin photocurrent caused by ${\bf k}$-linear terms in the
interband matrix elements has the form
\begin{equation}
J_{x'}^{z'} = \beta \varepsilon (e_{y'}^2-e_{x'}^2)
\frac{\tau_e}{\hbar} \frac{\eta_{cv}}{\hbar\omega} I \:, \;\;
J_{y'}^{z'} = \beta \varepsilon e_{x'} e_{y'} \frac{\tau_e}{\hbar}
\frac{\eta_{cv}}{\hbar\omega} I \:,
\end{equation}
where $\varepsilon=(\hbar\omega-E_g)m_h/(m_e+m_h)$ is the kinetic
energy of the photoexcited electrons, ${\bf e}=(e_{x'},e_{y'},0)$
is the light polarization vector, $y'\|[00\bar{1}]$. In contrast
to Eq.~(1), this contribution depends on the polarization plane of
the incident light and vanishes for unpolarized light. The spatial
separation of opposite spins, which depends on the light
polarization, has been observed in Ref.~\cite{Zhao}. However,
estimations show that the contributions~(1)~and~(3) in GaAs-based
QWs are comparable in magnitude for the excitation with 100~meV
above the band edge.

In (001)-grown QWs the absorption of linearly- or unpolarized
light results in a in-plane flow of electron spins. In contrast to
the low-symmetry QWs considered above, in (001)-QWs the
linear-in-{\bf k} terms in the matrix elements of optical
transitions under normal incidence vanish, and the spin
photocurrents are entirely related to the spin-orbit splitting of
the free-carrier subbands.

\paragraph*{Intrasubband optical transitions.} Light absorption by
free carriers, or the Drude-like absorption, is accompanied by
electron scattering by acoustic or optical phonons, static defects
etc. Scattering-assisted photoexcitation with unpolarized light
also gives rise to a pure spin current~\cite{TI,Ganichev}.
However, in contrast to the direct transitions considered above,
the spin splitting of the energy spectrum leads to no essential
contribution to the spin current induced by free-carrier
absorption. The more important contribution comes from asymmetry
of the electron spin-dependent scattering.

\section{Optical orientation by light of zero helicity}

Optical excitation with linearly polarized light in QWs can also
result in the spin orientation of photoexcited carriers. This
effect is related to the reduced symmetry of QW structures as
compared to bulk crystals and forbidden in bulk cubic
semiconductors. Theory of the optical orientation by linearly
polarized light has been developed in Refs.~\cite{oo1,oo2} for the
direct interband and indirect intrasubband transitions.
Microscopically, the effect is a two-stage process involving (i)
asymmetrical spin-dependent photoexcitation of the carriers
followed by (ii) spin precession in an effective magnetic field
induced by the Rashba or Dresselhaus spin-orbit coupling.
Direction of the average spin is determined by the structure
symmetry and geometry of photoexcitation. In (001)-grown QWs, the
linearly-polarized normal-incidence excitation results in the spin
orientation along the QW normal, with the spin sign and magnitude
depending on the orientation of light polarization plane.
Estimations show that the spin orientation by linearly polarized
light can reach a few percents and is experimentally accessible.

\paragraph*{Acknowledgments.}
This work was supported by the RFBR, RAS, President Grant for
young scientists, Russian Science Support Foundation, and
Foundation ``Dynasty'' - ICFPM.


\begin{thebibliography}{8}
\bibitem{Bhat} R.D.R.~Bhat, F.~Nastos, A.~Najmaie, and
J.E.~Sipe, {\em Phys. Rev. Lett.} {\bf 94}, 096603 (2005).
\bibitem{TI} S.A.~Tarasenko and E.L.~Ivchenko, {\em JETP Lett.}
{\bf 81}, 231 (2005).
\bibitem{Zhao} H.~Zhao, X.~Pan, A.L.~Smirl et al., {\em Phys. Rev.
B} {\bf 72}, 201302 (2005).
\bibitem{Ganichev} S.D.~Ganichev, V.V.~Bel'kov,
S.A.~Tarasenko et al., {\em Nature Physics} {\bf 2}, 609 (2006).
\bibitem{Khurgin} J.B.~Khurgin, {\em Phys. Rev. B} {\bf 73},
033317 (2006).
\bibitem{Jancu} J.-M.~Jancu, R.~Scholz, E.A.~de~Andrada~e~Silva,
and G.C.~La~Rocca, {\em Phys. Rev. B} {\bf 72}, 193201 (2005).
\bibitem{oo1} S.A.~Tarasenko, {\em Phys. Rev. B} {\bf 72}, 113302
(2005).
\bibitem{oo2} S.A.~Tarasenko, {\em Phys. Rev. B} {\bf 73}, 115317
(2006).
\end{thebibliography}
\end{document}